# Dual-task Coordination in Children and Adolescents with Attention Deficit Hyperactivity Disorder (ADHD)

Ketevan Inasaridze, Vera Bzhalava

Tbilisi State Medical University

**Abstract**

The deficit of executive functioning was found to be associated with attention deficit hyperactivity disorder (ADHD) in general and its subtypes. One of the important functions of central executive is the ability simultaneously coordinate two tasks. The study aimed at defining the dual-task performance characteristics in healthy children and adolescents on the computerised and the paper and pencil dual-task methods; investigating the effect of task difficulty on dual-task performance in ADHD in comparison to age and years of education matched healthy controls; testing if the paper and pencil version of the dual-task method is giving the same results in ADHD and healthy controls; investigating whether the dual-task functioning in ADHD is defined by the deficits in the general motor functioning and comorbidity factors. The study investigated dual task functioning in 6-16 years old 91 typically developing controls and 91 children with ADHD. It was found that: (1) the dual-task coordination is available in children and adolescents with ADHD in general and in its subtypes and not significantly different from performance of age and years of education matched healthy controls; (2) Increase of the task difficulty in dual-task paradigm don't affect disproportionately children and adolescents with ADHD in comparison to age and years of education matched healthy controls; (3) The paper and pencil version of the dual-task method is giving the same results in ADHD and healthy controls as computerised version; (4) The dual-task functioning in ADHD in general and in its subtypes is not defined by the general motor functioning while in healthy controls dual task performance is associated with the general motor functioning level; (5) The dual-task functioning in ADHD in general and in its subtypes is not defined by the comorbidity factors.

This research was supported by a grant N 021-08 from the Rustaveli National Science Foundation.

Correspondence concerning this article should be addressed to Ketevan Inasaridze
E-mail: kate_inasaridze@hotmail.com

Attention deficit hyperactivity disorder (ADHD) is one of the most prevalent developmental disorders characterized by developmentally inappropriate levels of inattention, impulsivity, and hyperactivity. ADHD makes serious difficulties for families, schools and society. ADHD is overwhelming public health problem with impact on economic and social life of many countries. Symptoms of ADHD often persist into young adulthood, and long-term consequences include lower educational and occupational achievement and increased risk for developing other psychiatric disorders (Mannuzza et al., 1997; Mannuzza et al., 1998).

According to the prevalent view on ADHD, it is a disorder of behavioral control in terms of inhibition. Barkley (1997a) proposed the one of the most widely cited theory of ADHD in which behavioral inhibition is regarded as a mechanism that develops early in life and underpins a number of other cognitive functions, among them executive functions, attention, working memory. As was proposed by the model of multiple developmental pathways (Sonuga-Barken EJS, 2005) ADHD is associated with the disturbances in motivational processes and executive dysfunction.

On the basis of several studies ADHD has been associated with deficits of executive functioning (Barkley, 1997b; Pennington et al., 1996; Willcutt et al., 2005; Bental & Tirosh, 2007). The different executive functions, measured by different neuropsychological tasks, showed that some of the executive processes are compromised in ADHD but others are not (Geurts et al., 2005; Passini et al., 2007). It was also found that ADHD subtypes can be characterized by different profiles of deficits in executive functioning and attention (Klorman et al., 1999; Lockwood et al., 2001; Nigg et al., 2002; Geurts et al., 2005).

According to the DSM-IV (APA, 1994) three subtypes of ADHD are defined: ADHD combined subtype (ADHD-C), ADHD predominantly hyperactive/impulsive subtype (ADHD-H) and ADHD predominantly inattentive subtype (ADHD-I). The validity of differentiating of these subtypes is rather debatable (Barkley et al., 1999; Milich et al., 2001). Some researchers view ADHD subtypes as different disorders with different cognitive, behavioral profiles and underlying neurobiologies (Diamond, 2005). According to Barkley's model of ADHD executive function deficit is associated with only ADHD-C and ADHD-H subtypes but not ADHD-I. There are controversial findings regarding distinction of the subtypes of ADHD. In some studies distinction between ADHD-H/Impulsive and ADHD-I were found (Milich et al., 2001; Geurts et al, 2005), but it was not found in other studies (Barkley et al., 1992; Faraone et al., 1998; Chhabildas et al., 2001; Murphy et al., 2001). In several studies was found that not only ADHD-C and ADHD-H subtypes are characterized by the deficit of executive functioning, but also ADHD-I and that profile of the deficits of executive functions are different for different subtypes of ADHD (Klorman et al., 1999; Lockwood et al., 2001; Nigg et al., 2002; Geurts et al., 2005). It is still needed to answer the question if the subtypes of the ADHD are distinctive diagnostic entities, or they don't differ from each other on neuropsychological measures. It is not clear cut which executive processes are compromised in each of them.

The study made contribution in understanding of the differences in cognitive characteristics of different types of ADHD.

Working memory is a complex concept with several possible definitions (Miyake & Shah, 1999). The presented study will rely on the working memory model proposed by Baddeley and Hitch (1974). This model of working memory has not been widely applied to ADHD, so studies of ADHD may provide on one hand some independent validation of it and on the other hand will gain our knowledge of cognitive dysfunction of ADHD and it subtypes.

The central executive is the less well studied component of working memory. One of the important functions of it is the ability simultaneously coordinate two tasks. As was shown in several studies the failure of this coordination is a characteristic impairment of patients with mild Alzheimer's disease both in a laboratory setting (Baddeley et al., 1986; Della Sala et al., 1995; Greene et al., 1995; Baddeley et al., 2001; MacPherson et al., 2004) and in everyday tasks (Alberoni, et al., 1992). Dual-task paradigm also proved to be the sensitive tool for detection of cognitive decrement in early stages of vascular dementia patients (Inasaridze et al., 2006a; Inasaridze et al., 2006b).

Moreover, a mild decrement in performance using paper and pencil version of the dual-task was shown in patients with Parkinson's disease (Dalrymple-Alford et al., 1994) and the test proved useful in differentiating patients with frontal lobe damage from patients with hippocampal damage (Cowey & Green, 1996). The dual-task appeared to be more sensitive than "frontal" tests to behavioral changes arising from frontal lobe damage. It proved better in differentiating brain damaged patients with lesions damaging the frontal lobes from those without a full-blown dysexecutive syndrome then the classic "frontal" tests like verbal fluency or the Wisconsin Card Sorting Test (Baddeley et al., 1997). The impairment on dual-task performance also was found in adults with autism (Garcia-Villamisar & Della Sala, 2002) that shows the importance of excluding participants with symptoms of autism from the dual-task studies of ADHD as we plan to do.

There are very few studies on dual-task performance in children and adolescents with ADHD. In different dual-task studies on ADHD different dual-task methods are used (West et al., 2002; Cornoldi et al., 2001; Leitner et al., 2007; Savage et al., 2006; Fuggetta, 2006; Wimmer et al., 1999; Karatekin, 2004; Schachar & Logan, 1990). In majority of them performance on both individual tasks was not adapted to the individual ability levels of each participant. Only in some studies titration procedure was used just for one task (Karatekin, 2004) but not for both of them. This makes difficult to find existence of dual-task deficit in children and adolescents with ADHD and its subtypes in comparison to healthy controls. Some of the individual tasks used in the previous studies in the dual-task paradigm are complex in respect of the load of other cognitive functions and the study results reflect a difficulty in dealing with those cognitive functions rather than a difficulty with dual-task performance. None of the previous studies investigated differences in dual-task performance of children and adolescents with ADHD in general and its subtypes using the dual-task paradigm when each subject perform each task on their individual level of ability that was the aim of the proposed study.

It is essential to investigate whether possible differences in dual-task performance in ADHD in general and in its subtypes are determined by task difficulty rather than by an overall problem with dual task coordination. The presented study investigated possible interaction between task demand and the need to divide attention (Logie et al., 2004).

In most studies a computerized version of the dual-task was used (Baddeley, 1986; Baddeley et al., 1991; Logie et al., 2004; MacPherson et al., 2004). The performance of computerized dual-task requires a light pen that is not a standard piece of equipment in most laboratories and computer programs to run the tasks, which may not readily transferable from one computer to the other. Therefore a paper and pencil version of the dual-task paradigm was devised (Baddeley et al., 1997; Della Sala et al., 1995). The paper and pencil version of the dual task was refined and the "Tbilisi paper and pencil motor task" was made in a user-friendly form for clinical settings on dementia patients (Inasaridze et al., 2006b). In the study the newly refined paper and pencil motor task was used for investigating the dual-task performance in healthy children and adolescents and in subjects with ADHD. It was tested whether paper and pencil version of the dual-task is successfully applicable in the educational settings.

The inconsistence found in results of the cognitive functioning in ADHD studies could be defined by different factors. One of them is the methodological differences. In several studies of executive functions, attention and working memory comorbid disorders are not controlled. Comorbid disorders like conduct disorder or oppositional defiant disorder, autism spectrum disorder, obsessive-compulsive disorders, Tourette syndrome, learning disability, are in itself associated with executive dysfunctions (Sergeant et al., 2002; Ozonoff & Strayer, 1997; Geurts et al., 2004). One of the problems that prevents to uncover the deficits in working memory functioning and in other cognitive functions in children and adolescents with ADHD in general and in its subtypes is the complex nature of the tasks that measure executive functions. Different researchers operationalize the same executive functions differently that resulted in using tasks which are heavily loaded not just on one executive function but several. In addition, performance of tasks on executive functions taps non-executive function cognitive processes that make difficult to uncover basic deficits in executive functioning. It makes unclear any existing differences in cognitive functioning in different subtypes of ADHD. The relatively large samples of ADHD and control subjects were taken to provide enough statistical power for the statistical analysis and to help uncovering of existing differences between the study groups. In the study subjects were carefully matched on age, years of education and IQ that also helped to reveal differences between the study groups.

The disorder of the development of motor functioning is found to be associated with the ADHD (Harvey & Reid, 2003; Dewey et al., 2007). It is necessary to control for general motor function deficits to study whether executive functioning deficit is specifically impaired in ADHD in general and its subtypes in comparison with the healthy controls. Thus in the study was investigated whether differences in dual-task performance are determined by the disorder in general motor functioning characterised to ADHD.

The study investigated if different subtypes of ADHD could have different profiles of deficits in executive functioning, particularly in dual-task coordination ability.

The study aimed at defining the dual-task performance characteristics in healthy children and adolescents on the computerised and the paper and pencil dual-task methods; determining developmental changes of the dual-task coordination in children and adolescents with ADHD in general and in its subtypes (ADHD-C, ADHD-H and ADHD-I) in comparison to age, years of education and intelligence level matched healthy controls; investigating whether any increase of the task difficulty in dual-task paradigm would disproportionately affect children and adolescents with ADHD in general and in its subtypes (ADHD-C, ADHD-H and ADHD-I) in comparison to age, years of education and intelligence level matched healthy controls; Testing if the paper and pencil version of the dual-task method is giving the same results in ADHD, its subtypes (ADHD-C, ADHD-H and ADHD-I) and healthy children and adolescents as computerised version of the dual task; Investigating whether the deficit in dual-task functioning in children and adolescents with ADHD in general and in its subtypes (ADHD-C, ADHD-H and ADHD-I) is defined by the deficits in the general motor functioning; Investigating whether the comorbidity factors (conduct disorder or oppositional defiant disorder) play a role in determining deficits of the dual-task performance in ADHD in general and in its subtypes (ADHD-C, ADHD-H and ADHD-I) in comparison to age, years of education and intelligence level matched healthy controls.

## Methods

### *Participants*

Case-control study of dual-task coordination was performed by forming groups of children and adolescents with ADHD and healthy children and adolescents. The healthy children and adolescents were included in the study for determining any developmental deviation in dual-task performance in children and adolescents with ADHD in comparison to healthy controls.

The study investigated dual-task functioning in 6-16 years old 91 typically developing controls and 91 children with ADHD. The healthy control children were recruited from different schools of Tbilisi. The children/adolescents with ADHD were recruited from different schools of Tbilisi and the Neurology Department of the Pediatric Clinic of the Tbilisi State Medical University. The 91 children with ADHD have been diagnosed in accordance with the DSM-IV criteria for ADHD, its subtypes (ADHD-C, ADHD-H and ADHD-I) and of comorbid disorders as conduct disorder (CD) or oppositional deviant disorder (ODD) supplemented by the information from the Parent/Teacher Disruptive Behavior Disorder (DBD) Rating Scale (Pelham et al., 1992). The normal controls (NC) had a mean of DBD 23.50 (SD=13.11) and children/adolescents with ADHD had a mean of DBD 64.25 (SD=20.17). The 91 NC children were matched as closely as possible to 91 children/adolescents with ADHD group for age and years of education. The healthy controls had a mean age of 9.65 (SD=2.66) and children/adolescents with ADHD had a mean age of 9.73 (SD=2.68). The NC had a mean years of education of 4.46 (SD=2.65) and ADHD group had a mean years of education of 4.46 (SD=2.66). There was no statistically significant difference between the total raw scores of WISC-III for NC (mean=269.77, SD=73.17) and ADHD groups (mean=252.01, SD=81.34), t(175)=1.525, p=.129. The significantly more boys with ADHD (64%) than NC boys (46%) participated in the study ($\chi^2$ = 5.682, df=1 p<.017). The age range is reported into three groups – 6-6.11 years, 7-10.11 years and 11-16 years, according to the MABC-2 age range division. Table 1 reports percentage distribution of sex, age, ADHD subtypes and comorbid disorders for healthy controls and children/adolescents with ADHD (Figure 1).

Table 1. The percentage distribution of sex, age, ADHD subtypes and comorbid disorders for controls and children/adolescents with ADHD.

| 182 subjects N (%) | | Control 91 (50%) | ADHD 91 (50%) | |
|---|---|---|---|---|
| Sex | Girls 82 (45.1%) | 49 (59.8%) | 33 (40.2%) | |
| | Boys 100 (54.9%) | 42 (42.0%) | 58 (58.0%) | |
| Age | 6-6.11 years 21 (11.5 %) | 10 (47.6%) | 11 (52.4%) | |
| | 7-10.11 years 98 (53.8 %) | 49 (50.0%) | 49 (50.0%) | |
| | 11-16 years 63 (34.6%) | 32 (50.8%) | 31 (49.2%) | |
| Comorbidity | | 9 (12.2 %) | 65 (87.8%) | |
| | ODD | 8 (12.1 %) | 58 (87.9%) | |
| | CD | 3 (6.5 %) | 43 (93.5%) | |
| ADHD subtypes | | Inattentive 21 (23.1 %) | Hyperactive/impulsive 22 (24.2 %) | Combined 48 (52.7 %) |

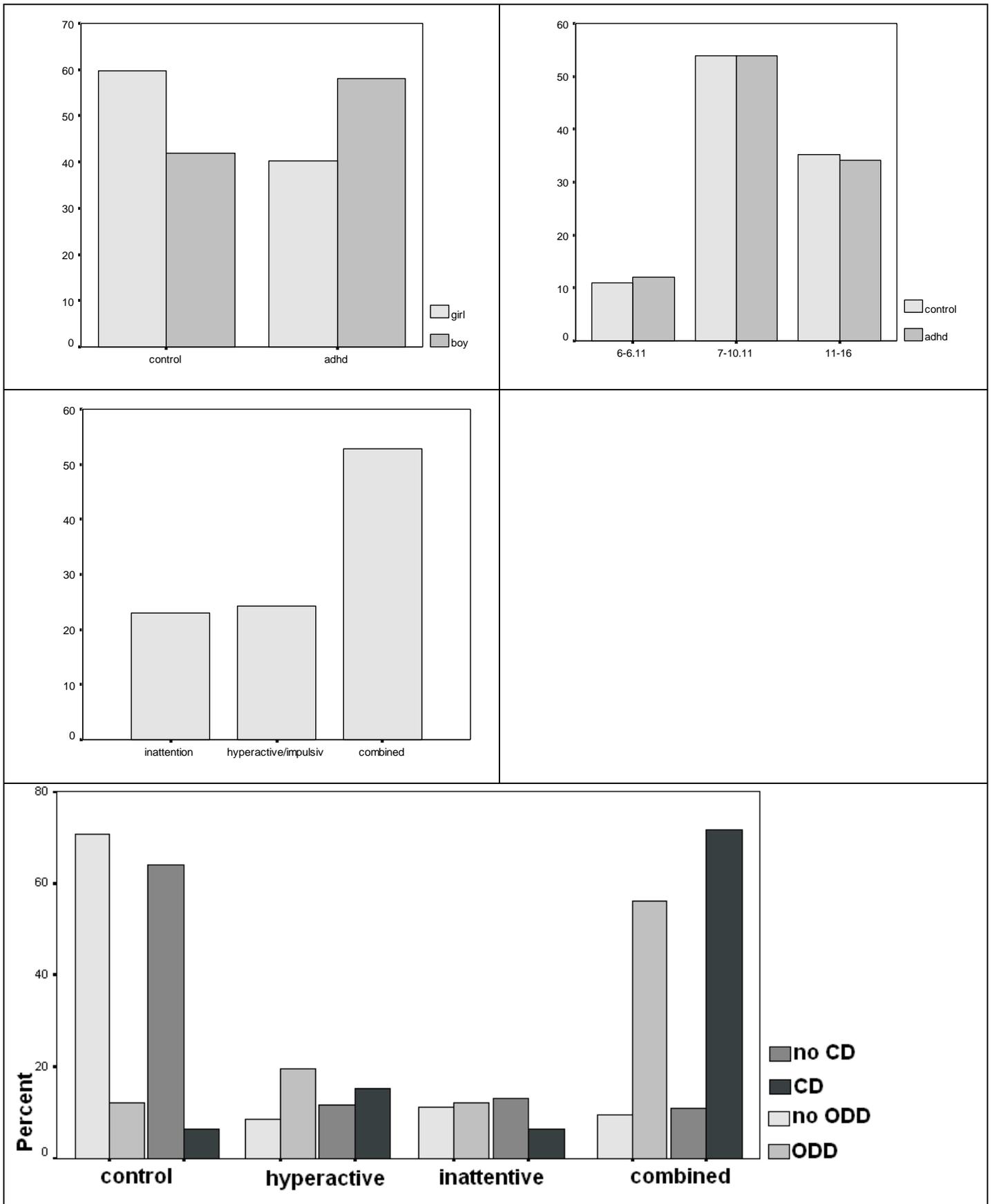

Figure 1. The percentage distribution of sex, age, ADHD subtypes and comorbid disorders for controls and children/adolescents with ADHD.

Additionally, the SNAP-IV Teacher and Parent Rating Scale (Swanson et al, 1983) was used to aid more clear determination of diagnosis of ADHD and confirmation differentiating of ADHD subtypes and assessment of the comorbid disorders. Symptoms of the autism spectrum disorder were assessed using the Social Communication Questionnaire (SCQ) (Berument et al., 1999). Children and adolescents with symptoms of Autism were excluded from the study. Parents/carers of all children and adolescents (controls and ADHD) were filled out checklist on developmental and medical history of children that include questions on pregnancy, delivery and infancy complications, early development, social and academic functioning. To make sure that the diagnoses was based on reports from the multiple informants familiar with the child's behavior in different settings parents/carers and teachers of the children and adolescents with ADHD rated children/adolescents on the general behavioral symptoms (Achenbach Child Behavior Checklist for Ages 6-18 (CBCL/6-18) and Achenbach Teacher Report Form (TRF/6-18), 2001) respectively. Children and adolescents with ADHD underwent general neurological and neuropsychological examination and subjects with history of mental retardation (full scale IQ less than 80 assessed by WISC-III), brain trauma or significant neurological conditions such as seizures for example, with physical impairment that prevent execution of the neuropsychological tasks, with severe problems of impressive and expressive language functions were excluded from the study. Potential control participants were excluded from the study if they had attention problems in the past for which they were seeking help or if they had any first-degree family members diagnosed with ADHD.

The subjects with 4 or 5 symptoms on either inattention or hyperactive/impulsive or both dimensions were excluded from the analyses to overcome inclusion of borderline cases and leave for the analyses more expressed forms of ADHD and healthy controls with less symptoms of inattention or hyperactivity/impulsivity.

The inclusion of no less than 30 participants in each study group (control or ADHD) ensured enough statistical power for data analyses.

*Apparatus and Material*

General intelligence level of participants was assessed by the WISC-III. The DBD (Pelham et al., 1992) was used to help determining if a child/adolescent meets the criteria for DSM IV diagnoses of ADHD and its subtypes (ADHD-C, ADHD-H and ADHD-I), ODD, or CD. The clear determination of diagnosis of ADHD, additional confirmation of its subtypes and assessment of the comorbid disorders were performed by the SNAP-IV (Swanson et al, 1983). Symptoms of the autism spectrum disorder was assessed by the SCQ (Berument et al., 1999). The general behavioral symptoms of children and adolescents were assessed by parents/carers and teachers using Achenbach CBCL/6-18 and TRF/6-18 (2001) respectively. The profile of deficits in general motor functioning was be assessed by the Movement Assessment Battery for Children – Second Edition (MABC-2) (Henderson & Sugden, 2007).

Information about developmental and medical history of children/adolescents that include questions on pregnancy, delivery and infancy complications, early development, social and academic functioning was collected from parents/carers using checklist adapted in Memory Clinic, Tbilisi (2006) that is based on modified questionnaire from Ehrenberg MF (Spreen & Strauss, 1998).

Children and adolescents with ADHD were underwent general neurological and neuropsychological examination to exclude from the study subjects with brain trauma or significant neurological conditions such as seizures for example, with physical impairment that prevent execution of the neuropsychological tasks, with severe problems of impressive and expressive language functions.

The computerised dual-task experiments were run on a Pentium 2 (monitor screen 314mm×216mm). The light-pen hardware was used for the computerized motor tracking tasks. The experimental material for the dual-task experiment were include computerized and paper and pencil ("Tbilisi paper and pencil motor task") versions (Inasaridze et al., 2006b) of the dual-task (digit span - motor tracking).

The neurological and neuropsychological assessments were carried out by the trained neurologist and qualified neuropsychologist. The interviews and scales were conducted by the researchers trained in clinical psychology.

**The computerized version of the dual-task**

In the computerised version of the dual-task study participants were performed the List Memory Task – serial digit recall verbal task and a computerised version of the perceptuomotor tracking task singly and in a dual-tasks paradigm whereby the two individual tasks are performed simultaneously.

*List Memory Task*. Participants listened to lists of digits from a computer and repeated the digits in serial order. All nine digits (1-9) were recorded by a professional TV speaker and after using computer program Cool Edit Pro 2.0 and Superlab 1.03 were randomly combined in lists of digits of different length. In each list digits were presented at a rate of 1 per second. In the Digit Span Determination participants were tested on six lists of the same length, starting with length 2. Participants' digit span was determined as the maximum length of the lists of which the participants recalled at least 5/6 correctly. In the List Memory Single Task each subject immediately repeated back the lists, the length of which was equal to the subjects span during 2 minutes. Two different scoring procedures were used for the final score of the List Memory Task. According to the rule of scoring number of correctly recalled lists was divided by the number of lists presented (Cocchini et al., 2002).

*Motor Tracking Task*. A target comprising a red oval with dark spots about 2.5 cm long and 2 cm wide was shown on a computer screen to the study subjects. This stimulus is resembles an insect known as a "ladybug" in North America and known as a "ladybird" in the UK. Subjects were given a light-sensitive stylus that they placed on the target which then began to move randomly around the screen. The task requires keeping the stylus placed on the moving Ladybug the speed of which could be set at different levels. At the speed level 1 the target was moved approx. 3.5 cm per second. Different speed levels of the target differed from each other by 1 cm per second (speed level 2 – 4.5 cm per second, speed level 10 – 12.5 cm per second). The target remained red as long as the stylus was in contact, but changed to green when contact was lost and returning to red when contact was regained. The monitor screen was placed at an angle of 30 degrees from horizontal, because in previous studies it was found that this angle is less physically tiring than using a vertical screen. In the adaptive tracking test the Ladybug moved slowly at speed level 2 = 4.5 cm per second. The speed level increased to the next level if, over a period of 5 seconds, the participant maintained contact with the target for at least 60 % of the time. If time on target was less than 40 %, the speed level was reduced to the level below. If time on target was between 40 % and 60 %, the speed level did not change. When the speed level remained constant for 15 seconds – three 5 second periods, the adaptive tracking phase was complete, and this speed level was used as a measure of the tracking ability for the individual. To avoid fatigue from a lengthy adaptive tracking phase, speed level changes at the lower level involved single steps from 1 to 5, whereas higher speed levels involved changes of two steps at any given time. This also allowed for lower ability people to have a reasonable amount of practice, but at the same time higher ability subjects would not perform a dramatically larger number of trials, which would possibly result in substantially different levels of practice according to individual ability (Cocchini et al., 2002).

*The Dual-task condition*. In the dual-task condition subjects performed the tracking task while simultaneously verbally reproducing lists of digits.

**The paper and pencil version of the dual-task**

In the paper and pencil version of the dual-task study subjects were performed the List Memory Task – serial digit recall verbal task and paper and pencil version of the perceptuomotor tracking task singly and in a dual-tasks paradigm whereby the two individual tasks are performed simultaneously.

*List Memory Task*. The same serial digit recall task used for the computerised version of the dual-task was used for the paper and pencil version of the dual-task.

*The "Tbilisi paper and pencil motor task"*. In the paper and pencil version of the motor tracking task filled black arrows were arranged in vertical parallel lines. The filled black arrows prevent writing of numbers inside the arrows. Arrows also indicated direction to move on. Participants were presented with 373 black arrows linked with each other and forming path laid out on an A3-size sheet of white paper (Figure 2).

Height of the tip of each arrow was 5 mm and length of each base was 7 mm. Straight lines linking arrows were 1 cm in length. Subjects were required to use a felt pen to cross out arrows from the start arrow to the end, as was indicated on the paper. They place a cross on each successive arrow as quickly as possible. Number of arrows was chosen in pilot trials so that it was impossible to cross all arrows in two minutes. Participants were first given a number of practice trials with a short, 35-arrow path presented on an A4-size paper, to accustom them to the procedure, and to ensure that they understood task requirements. The score of the motor task was the number of arrows successfully marked by the participant.

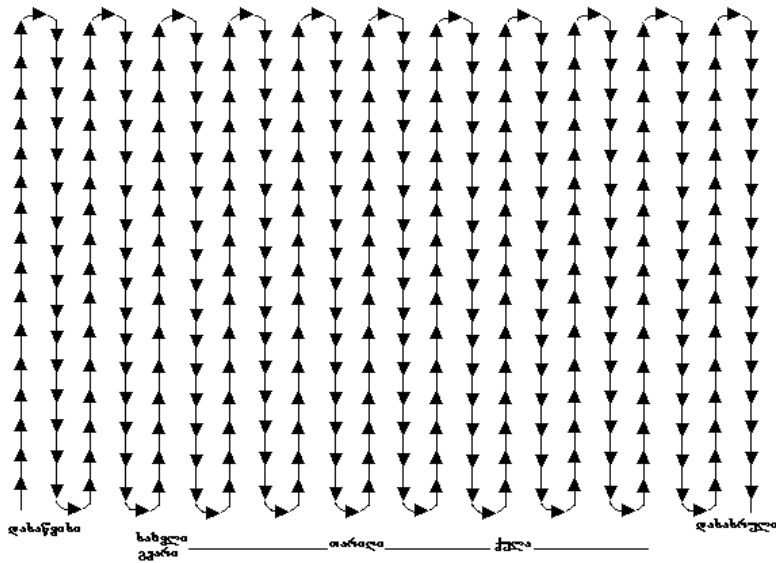

Figure 2. Pattern of arrows for motor tracking task on the "Tbilisi paper and pencil motor task".

*The Dual-task condition*. In the dual-task condition participants performed the tracking task while simultaneously verbally reproducing the lists of digits.

## Procedure

The study was conducted in agreement with Helsinki Declaration (1977) regarding ethical standards for clinical studies in medicine. Written informed consents were obtained from parents/careers of children/adolescents. Study included only those children/adolescents whose parents/careers agreed to take part in the research. Experiments presented in this study were discussed and received approval from the Ethic Commission of the Tbilisi State Medical University.

Teachers from different schools of Tbilisi were underwent the preliminary interview to make them accustomed with the ADHD symptoms and found children/adolescents with the possible ADHD. Then the more thorough neurolopsychological and neurological examination of child/adolescent and teachers and parents/careres ratings of the child were performed. At first teacher and parent/career filled in the DBD. If the criterion of the ADHD was met, then the SNAP-IV and SCQ were administered. The comorbid disorders were assessed and children/adolescents with symptoms of autism were excluded from the study. Parents/carers of all children and adolescents (controls and ADHD) filled out checklist on developmental and medical history of children that include questions on pregnancy, delivery and infancy complications, early development, social and academic functioning. The general behavioral symptoms of children and adolescents were assessed by parents/carers and teachers using CBCL/6-18 and Achenbach TRF/6-18 (2001) respectively. The same procedure was used for the parents of ADHD children recruited from the Neurology Department of the Pediatric Clinic of the Tbilisi State Medical University.

Parents of the children/adolescents with the ADHD were asked to refrain from giving psychostimulants (but not other medication children taking on a daily basis) 24 hours prior of testing sessions and were warned about this 2 days before the testing session.

All children were administered WISC-III. Handedness of all participants was measured by tasks devised based on the Luria assessment methods.

Children and adolescents with ADHD were underwent general neurological and neuropsychological examination to exclude from the study subjects with history of mental retardation (full scale IQ less than 80 assessed by WISC-III), brain trauma or significant neurological conditions such as seizures for example, with physical impairment that prevent execution of the neuropsychological tasks, with severe problems of impressive and expressive language functions. The profile of deficits in general motor functioning of children/adolescents with ADHD was assessed by the MABC-2.

The **dual-task experiments** were conducted on one experimental session. In the beginning of experiments the Digit Span Determination procedure was performed for all participants.

In dual-task experimental session, using the computerized version of the dual-task, participants performed the List Memory (single task), the Computerized Tracking Task (single Task), then computerized dual-task (List Memory plus tracking), the retest of the dual-task condition and lastly the List Memory (single task) and the Computerized Tracking Task (single Task) were administered again by the same experimenter in order to account for practice, motivation and fatigue effects. Each single task as well as the dual task was last two minutes. The presentation order of List Memory and tracking performed as single tasks were counterbalanced across participants. The presentation order of single and dual tasks were counterbalanced across participants.

In dual-task experimental session, using the paper and pencil version of the dual-task, participants performed the List Memory (single task), the Tbilisi Paper and Pencil Tracking Task (single Task), then paper and pencil dual-task (List Memory plus tracking), the retest of the dual-task condition and lastly the List Memory (single task) and the Tbilisi Paper and Pencil Tracking Task (single Task) were administered again by the same experimenter in order to account for practice, motivation and fatigue effects. Each single task as well as the dual task lasted two minutes. The presentation order of List Memory and tracking performed as single tasks were counterbalanced across participants. The presentation order of single and dual tasks were counterbalanced across participants.

The presentation order of computerized and paper and pencil versions of the dual-task were counterbalanced across participants.

To investigate the task difficulty effect on ADHD and its subtypes in dual-task paradigm three difficulty levels were defined for the List Memory (single task) and the Computerized Tracking Task (single Task). For the tracking task the low difficulty level was defined as 0.5 proportion of subject's individual tracking speed, standard level of difficulty was defined as subject's individual tracking speed and high level of difficulty as 1.5 proportion of subject's individual tracking speed. For the digit recall task the low difficulty level was defined as Span-2, standard level of difficulty was defined as subject's individual Span and high level of difficulty as Span+2. Participants performed each 6 single tasks (3 motor tracking and 3 digits recall tasks). The order of presentation of 6 single tasks was counterbalanced across participants. Then 6 dual-tasks were performed in 3 of which motor tracking task was at standard level and digit recall task varied in difficulty and in 3 of other dual-tasks digit recall task difficulty was fixed at span and the motor tracking task difficulty varied. The order of presentation of 6 dual-tasks was counterbalanced across participants. Each 12 single and dual tasks lasted two minutes (Logie et al, 2004).

All families of ADHD and control subjects were provided verbal and written feedback on the results of testing and/or diagnostic evaluation.

## Results

Data were analyzed by different descriptive and inferential statistical methods using SPSS 10.0.

The 2 groups × 2 conditions repeated measures ANOVA was used to determine the effect of dual task on performance of List Memory Task and motor tracking. Post hoc comparisons were determined by the Bonferroni test. The correlation analysis was made by the Pearson correlations. The independent-sample t test was used to make the between group comparisons. The regression models was tested by the stepwise backward conditional multiple regression method. The nonparametric Mann Whitney U test was used to make between group comparisons for independent samples.

**The computerized version of the dual-task.**

Table 2 reports the means of the digit span, List Memory Task and computerized motor tracking for healthy controls and children and adolescents with ADHD.

Table 2. Mean performances of the List Memory Task and motor tracking on computerized version of the dual task for study groups

| M (σ) | Digit span | List Memory Task | | Retest of List Memory Task | | Motor Tracking (% accuracy score) | | Retest of Motor Tracking (% accuracy score) | |
|---|---|---|---|---|---|---|---|---|---|
| | | Single | Dual | Dual | Single | Single | Dual | Dual | Single |
| Control | 4.45 (.69) | .82 (.12) | .78 (.13) | .78 (.18) | .80 (.14) | 54.26 (8.45) | 57.84 (9.72) | 57.07 (9.70) | 57.55 (9.87) |
| ADHD | 4.40 (.91) | .82 (.14) | .76 (.18) | .77 (.17) | .80 (.14) | 52.13 (9.13) | 53.67 (10.55) | 52.09 (11.27) | 53.69 (11.51) |

To determine the effect of dual task on performance of List Memory Task and motor tracking, the data from the single and dual tasks were entered separately into a 2 (group) × 2 (condition – type of task: single vs. dual) analysis of variance (ANOVA). For the computerised motor tracking task the ANOVA showed a significant effect of type of task, $F(1,179)=27.187$, $MSE=21.776$, $p<.0001$, $\eta^2=.132$; of group $F(1,179)=5.642$, $MSE=158.564$, $p<.018$, $\eta^2=.031$ and an interaction $F(1,179)=4.298$, $MSE=21.776$, $p<.04$, $\eta^2=.023$. For the List Memory Task the ANOVA showed a significant effect of type of task $F(1,179)=22.88$, $MSE=.009$, $p<.0001$, $\eta^2=.113$ but no effect of group ($F<1$) and no interaction ($F<1$). The control group showed significant improvement in computerized motor tracking task performance under the dual task condition compared to the single task performance than ADHD group subjects (Figure 2).

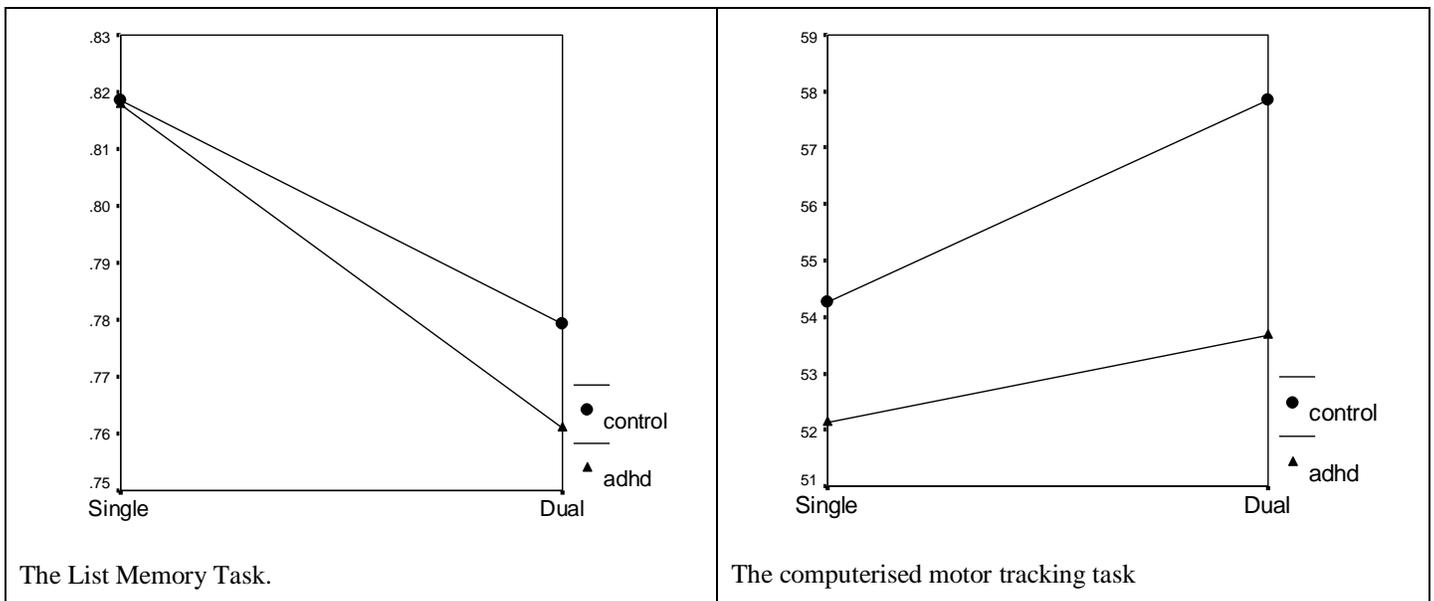

The List Memory Task. The computerised motor tracking task

Figure 2. The List Memory and computerised motor tracking task performance by healthy controls and children and adolescents with ADHD.

In the previous dual-task study (Logie et al., 2004) it was shown that reporting of the patterns for each individual task under dual task condition might be misleading, because this cannot account for the overall changes in performance across both tasks or for trade-offs in performance between tasks. Thus, an overall measure of performance – percentage change, was calculated for each participant. The percentage change combines the percentage change in accuracy that occurs between the single and dual tasks for the List Memory Task and the motor tracking task (Baddeley & Della Sala, 1996).

The percentage change formula is the following:

$$\text{Percentage change} = \frac{\text{Single task performance} - \text{dual task performance}}{\text{Single task performance}} \times 100$$

The percentage change for each task was combined as follows:

$$\text{Combined percentage change } (mu) = 100 - \frac{\text{Percentage change verbal} + \text{Percentage change tracking}}{2}$$

When this formula was applied to the validating study, a clear separation between performance of AD patients and control subjects was found (Baddeley & Della Sala, 1996).

Two percentage change scores were calculated for the List Memory Task and for the perceptuomotor tracking task. The μ score was determined for the computerized version of the dual-task. Table 3 presents these two percentage change scores and the μ score.

Table 3. The percentage change scores for the computerised version of the dual-task.

| M (σ)   | List Memory Task | Motor Tracking | Combined μ score |
|---------|------------------|----------------|------------------|
| Control | 4.06 (14.98)     | -7.07 (12.3)   | 101.51 (10.34)   |
| ADHD    | 6.51 (18.89)     | -3.36 (13.67)  | 98.43 (12.01)    |

The percentage change score on computerised motor tracking task was higher for the control group (mean=-7.07, SD=12.35) compared to the ADHD group (mean=-3.36, SD=13.67) but was significant at one tail $t(179)=-1.912$, $p<.057$. The μ score for control group (mean=101.51, SD=10.34) was higher in comparison to the group with ADHD (mean=98.43, SD=12.01), but was significant at one tail $t(179)=1.847$, $p<.066$ (Figure 3).

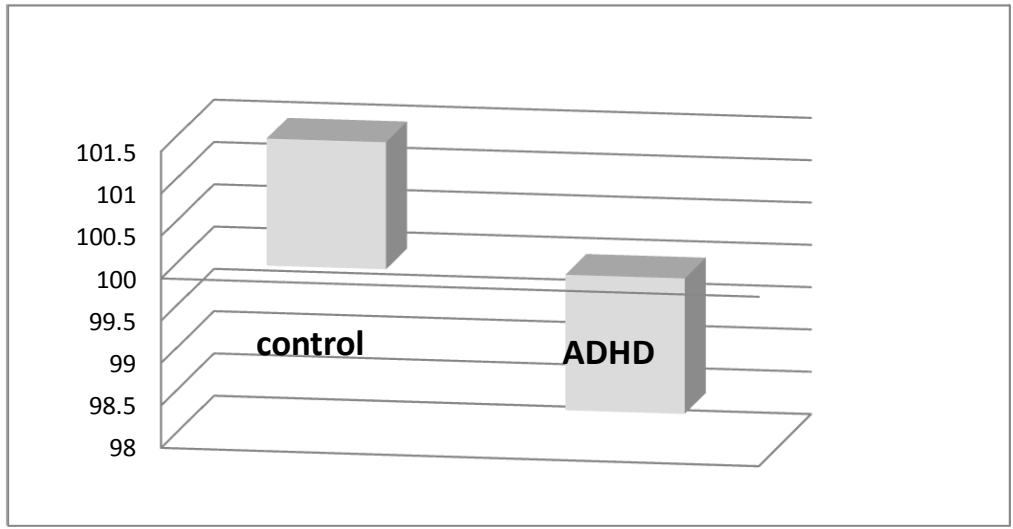

Figure 3. The combined percentage change µ scores for computerised version of the dual-task.

**The paper and pencil version of the dual-task.**

Table 4 reports the means of the digit span, List Memory Task and paper and pencil motor tracking for healthy controls and children and adolescents with ADHD.

Table 4. Mean performances of the List Memory Task and motor tracking on paper and pencil version of the dual task for study groups.

| M (σ) | Digit span | List Memory Task | | Retest of List Memory Task | Motor Tracking (% accuracy score) | | Retest of Motor Tracking (% accuracy score) | |
|---|---|---|---|---|---|---|---|---|
| | | Single | Dual | Dual | Single | Dual | Dual | Single |
| Control | 4.45 (.69) | .82 (.12) | .84 (.14) | .84 (.15) | 114.38 (37.71) | 113.96 (40.92) | 117.79 (38.47) | 123.51 (39.66) |
| ADHD | 4.40 (.91) | .82 (.14) | .80 (.14) | .81 (.15) | 111.23 (38.53) | 108.98 (40.07) | 111.72 (38.55) | 124.33 (39.99) |

For the paper and pencil motor tracking task and the List Memory Task the ANOVA didn't showed a significant effect of type of task, of group or interaction (Figure 4). The performance of healthy control subjects was better that of children/adolescents with ADHD but it didn't reach significance.

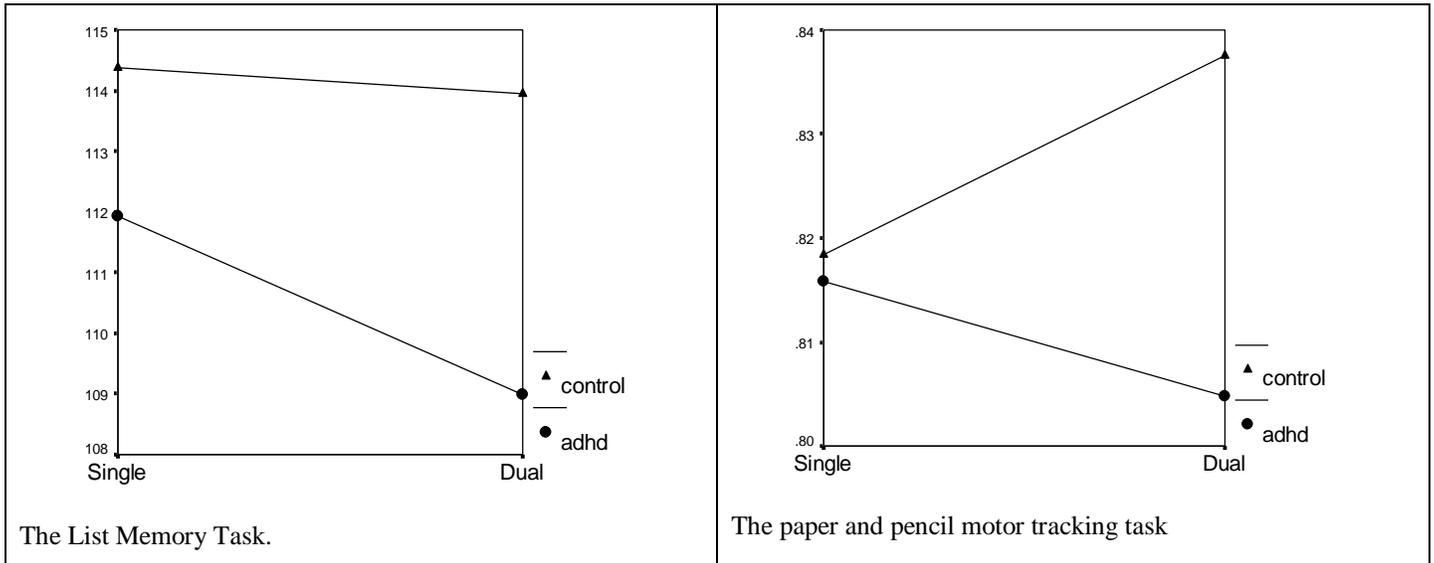

The List Memory Task.

The paper and pencil motor tracking task

Figure 4. The List Memory and paper and pencil motor tracking task performance by healthy controls and children and adolescents with ADHD.

Table 5 presents these two percentage change scores and the μ score for the paper and pencil version of the dual-task.

Table 5. The percentage change scores for the computerised version of the dual-task.

| M (σ) | List Memory Task | Motor Tracking | Combined μ score |
|---|---|---|---|
| Control | -3.27 (17.29) | .67 (13.12) | 101.21 (9.97) |
| ADHD | .08 (17.20) | 1.60 (18.61) | 99.14 (13.02) |

There was not found significant differences between percentage change and μ scores on paper and pencil version of the dual task for the study groups (Figure 5).

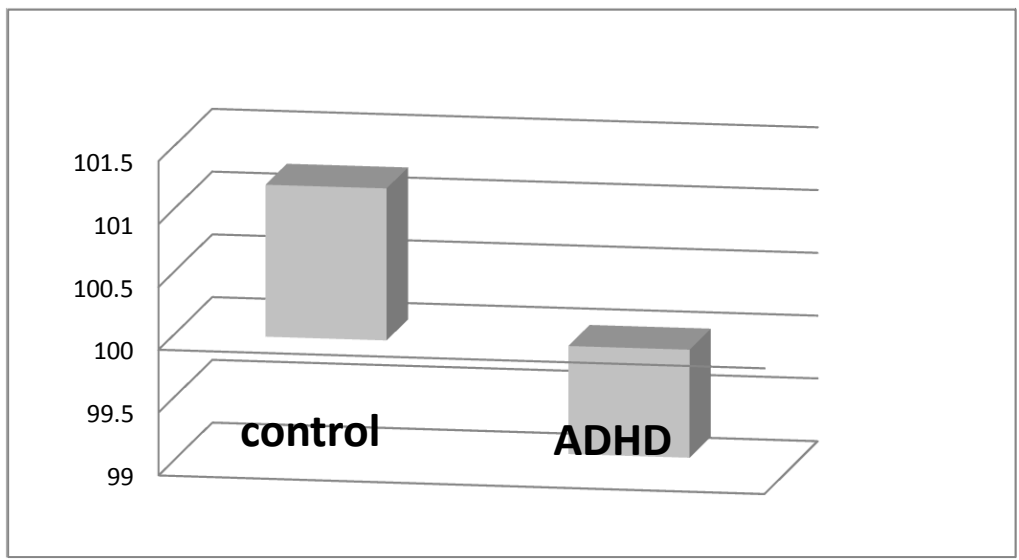

Figure 5. The combined percentage change μ scores for paper and pencil version of the dual-task.

**The correlation of the test-retest and of the computerised - paper and pencil versions of the dual-task.**

The moderate magnitude significant test-retest reliability was found for the computerised dual-task μ scores (r=.36, p<.0001) and for the paper and pencil dual-task μ scores (r=.37, p<.0001). For the healthy control group was revealed moderate magnitude significant test-retest reliability for the computerised μ scores (r=.50, p<.0001) and for the paper and pencil μ scores (r=.39, p<.0001). For the ADHD group the slightly low magnitude but significant test-retest reliability was found for the computerised μ scores (r=.24, p<.025) and for the paper and pencil μ scores (r=.33, p<.001).

The correlation between the μ scores of the computerised and paper and pencil versions of the dual-task was defined (r=.26, p<.0001) to find out whether these two versions of the dual task are comparable counterparts of each other. The moderate significant correlations were found for the group of healthy control participants (r=.29, p<.007) and group of children with ADHD (r=.23, p<.031).

**Influence of general motor functioning and comorbidity factors on dual-task coordination.**

Table 6 reports the means of subscales of the MABC-2 – the Manual Dexterity, Aiming & Catching, Balance and total score on MABC-2 for healthy controls and children and adolescents with ADHD.

Table 6. Mean performances on the MABC-2 scales and total performance for study groups.

| M (σ) | Manual Dexterity | Aiming & Catching | Balance | MABC-2 Total Score |
|---|---|---|---|---|
| Control | 84.30 (21.72) | 15.14 (5.53) | 63.95 (17.91) | 163.20 (24.38) |
| ADHD | 85.05 (16.83) | 14.57 (5.86) | 56.98 (16.04) | 156.59 (21.03) |

The general motor functioning assessed by the MABC-2 (Henderson & Sugden, 2007) revealed significantly better balance t(170.499)=2.713, p<.007 in healthy controls (mean=63.95, SD=17.91) compared to group with ADHD (mean=56.98, SD=16.04) but difference on the total test score was significant at one tail t(168.776)=1.918, p<.057 for the study groups.

For the computerised motor tracking task the 2 (group) × 2 (condition – type of task: single vs. dual) ANOVA showed a significant effect of interaction of type of task with the balance score F(1,171)=5.109, *MSE*=21.579, p<.025 taken as a covariate variable. For the List Memory Task the ANOVA showed a significant effect of interaction of type of task with the balance score F(1,171)=6.358, *MSE*=.009, p<.013 and the MABC-2 total score (1,171)=6.965, *MSE*=.009, p<.009 taken as a covariate variables. According to these results it can be suggested that high level of motor performance is associated with the less change in performance when subject moves from single task performance to dual task performance. The balance score was significantly correlated with the percentage change scores for the List Memory Task r=-.28, p<.01 and for motor tracking task r=-.22, p<.04. For the percentage change scores of the List Memory Task the Univariate Analysis of Variance showed a significant effect of balance (1,169)=6.128, *MSE*=269.257, p<.014 and MABC-2 total score (1,169)=4.98, *MSE*=271.033, p<.027 as covariate variables. For the percentage change scores of the motor tracking task the Univariate Analysis of Variance showed a significant effect of balance (1,169)=4599, *MSE*=159.433, p<.033 as covariate variable.

The computerised μ score for group of healthy controls was significantly correlated with the MABC-2 total score r=.26, p<.017, balance score r=.34, p<.002. For the computerised μ scores of healthy controls the Univariate Analysis of Variance showed a significant effect of balance (1,83)=9.521, *MSE*=95.042, p<.003 and MABC-2 total score (1,83)=5.71, *MSE*=99.124, p<.019 as covariate variables. For the ADHD group was not found significant correlations between computerised μ score and MABC-2 total score, or balance score.

The stepwise backward conditional multiple regression method showed that dependent variable the computerized μ score for healthy control group was significantly predicted by the balance score (standardised B=.22, p<.046) and WISC-III raw score (standardised B=-.28, p<.013) but for the group of children and adolescents with ADHD it was predicted by sex (standardised B=-.26, p<.015).

The Univariate Analysis of Variance for the μ scores didn't show significant effect of sex for the group of healthy controls but for the group of children/adolescents with ADHD was found significant effect of sex F(1,85)=5.194, *MSE*=122.063, p<.025. The boys with ADHD had significantly low mean μ score than healthy control boys (Figure 6). It wasn't found significant interaction between sex and group on μ scores.

The Univariate Analysis of Variance for the μ scores showed significant effect of balance F(1,77)=11.172, *MSE*=89.506, p<.001 and MABC-2 total score F(1,77)=5.098, *MSE*=96.128, p<.027 for girls but not for boys.

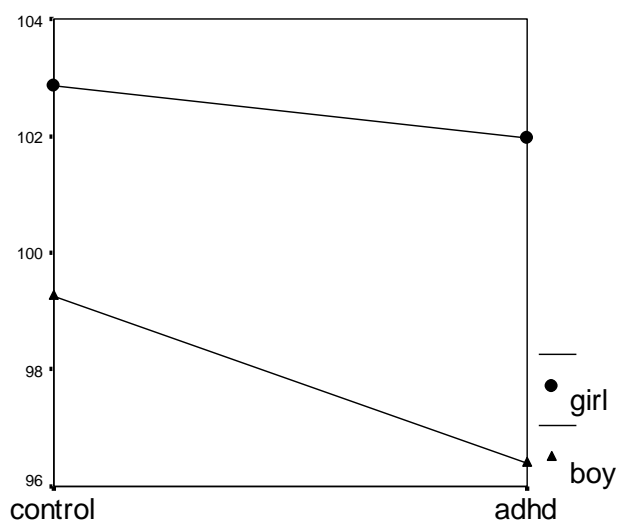

Figure 6. The means of the computerised μ scores for sex groups of healthy controls and children/adolescents with ADHD.

The paper and pencil μ score wasn't correlated with MABC-2 scale factors and total score for groups of healthy controls and children/adolescents with ADHD. There were no significant effects of the balance, MABC-2 total scores and sex on paper and pencil μ score for two study groups.

For the paper and pencil motor tracking task the 2 (group) × 2 (condition – type of task: single vs. dual) ANOVA showed a significant effect of interaction of type of task with the balance score $F(1,171)=11.953$, $MSE=2654.675$, $p<.001$ and the MABC-2 total score $(1,171)=4.851$, $MSE=2761.887$, $p<.029$ taken as a covariate variables. For the List Memory Task the ANOVA showed a significant effect of interaction of type of task with the MABC-2 total score $(1,171)=3.971$, $MSE=.008$, $p<.048$ taken as a covariate variable. It was found small negative correlations between balance score and single motor task $r=-.39$, $p<.0001$ and dual motor task $r=-.37$, $p<.0001$ and between the MABC-2 total score and single motor task $r=-.31$, $p<.003$ and dual motor task $r=-.27$, $p<.013$ for groups of healthy controls but not for ADHD group. The Univariate Analysis of Variance for the single motor task showed significant effect of balance $(1,170)=12.542$, $MSE=1317.774$, $p<.001$ and MABC-2 total score $(1,170)=4.885$, $MSE=1375.469$, $p<.028$ and for the dual motor task showed significant effect of balance $(1,169)=9.624$, $MSE=1497.677$, $p<.002$ and MABC-2 total score $(1,169)=4.398$, $MSE=1542.821$, $p<.037$.

The Univariate Analysis of Variance for the μ scores didn't show significant effect of sex, study group, balance and MABC-2 total score and their interaction for paper and pencil version of the dual-task.

The stepwise backward conditional multiple regression method showed that dependent variable the computerized μ score for healthy control group was significantly predicted by the manual dexterity score (standardised $B=.37$, $p<.004$), balance score (standardised $B=.26$, $p<.042$) and sex (standardised $B=.21$, $p<.089$) but none of variables significantly predict the μ score for the group of children and adolescents with ADHD.

To determine the effect of ADHD and sex on dual task for performance of computerized motor tracking, the data from the single and dual tasks were entered separately into a 2 (group: control vs. ADHD) × 2 (group: girls vs. boys) × 2 (condition – type of task: single vs. dual) analysis of variance (ANOVA). For the computerised motor tracking task the ANOVA showed a significant effect of sex by factor interaction $F(1,177)=11.011$, $MSE=20.713$, $p<.001$, significant effect of factor $F(1,177)=32.144$, $MSE=20.713$, $p<.0001$ and group (control vs. ADHD) $F(1,177)=5.506$, $MSE=160.34$, $p<.02$. In general in both groups, girls perform better than boys but in ADHD group boys didn't show improvement in dual-task condition as was found in healthy girls and boys and girls with ADHD (Figure 7).

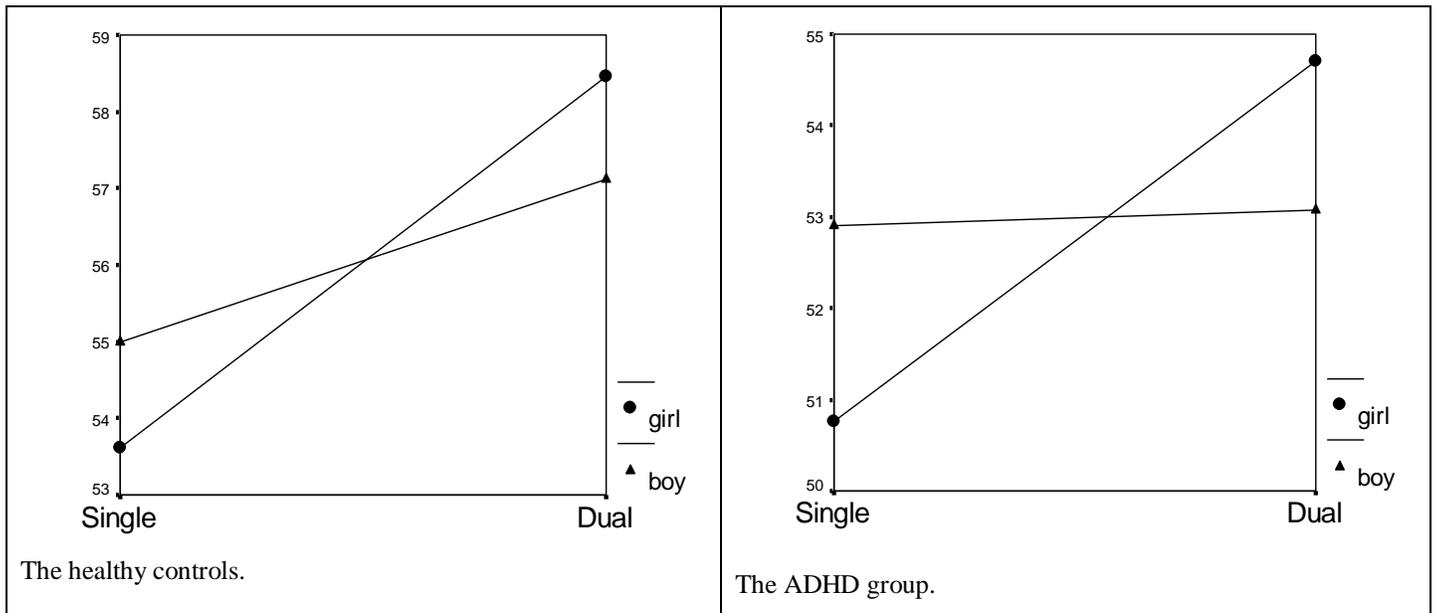

Figure 7. The mean performance of computerized motor tracking task by sex for the study groups.

To determine the effect of ADHD type and sex on dual task for performance of List Memory Task and computerised motor tracking, the data from the single and dual tasks were entered separately into a 4 (groups: control, inattentive, hyperactive/impulsive, combined) × 2 (group: girls vs. boys) × 2 (condition – type of task: single vs. dual) analysis of variance (ANOVA). For the computerised motor tracking task the ANOVA showed a significant effect of factor $F(1,85)=5.9$, $MSE=21.231$, $p<.017$ and of factor by sex interaction $F(1,85)=5.181$, $MSE=21.231$, $p<.025$. The effect of ADHD type, factor by ADHD types, of groups and 3 ways interaction were not significant. When the 4 (groups: control, inattentive, hyperactive/impulsive, combined) × 2 (group: with comorbidity vs. without comorbidity) × 2 (condition – type of task: single vs. dual) ANOVA was used it was found the significant effect of factor $F(1,173)=10.277$, $MSE=21.904$, $p<.002$ and of factor by ADHD types interaction $F(1,173)=2.955$, $MSE=21.904$, $p<.034$, of group (with comorbidity vs. without comorbidity) $F(1,173)=4.201$, $MSE=148.338$, $p<.042$ and of group (with comorbidity vs. without comorbidity) by group (controls and types of ADHD) interaction $F(3,173)=5.156$, $MSE=148.338$, $p<.002$. The group (controls and types of ADHD) effect was significant at one tail $F(3,173)=2.302$, $MSE=148.338$, $p<.079$. When sex, comorbidity and ADHD types were taken into account it was found significant effect of this 3 way interaction $F(1,165)=6.637$, $MSE=19.992$, $p<.011$. The nonparametric Mann Whitney U test was used to compare groups with and without comorbidity disorders for controls and ADHD subtypes. The nonparametric Mann Whitney U test showed the significant differences on computerised single motor task (for group with comorbidity $M=48.96$, $\sigma=6.12$; for group without comorbidity $M=63.23$, $\sigma=9.99$) $U=11$, exact $p<.001$ and dual motor task (for group with comorbidity $M=49.22$, $\sigma=7.92$; for group without comorbidity $M=62.05$, $\sigma=6.84$) $U=13$, exact $p<.001$ for ADHD group with hyperactivity/impulsivity and for single motor task (for group with comorbidity M=46.84, σ=5.36; for group without comorbidity M=53.67, σ=6.78) U=20, exact p<.014 ADHD group with inattention. In all cases groups with comorbidity showed the lower performance in comparison to groups without comorbidity.

The different repeated measures ANOVA conducted on the List Memory Task for computerised dual-task method, showed that the effect of the group (controls and types of ADHD) by comorbidity interaction F(3,165)=2.549, *MSE*=.03, p<.058 and of the group (controls and types of ADHD) by sex by comorbidity interaction F(3,165)=2.548, *MSE*=.03, p<.058 were significant at one tail.

The general motor functioning was not significantly different for the ADHD subtypes.

To determine the effect of ADHD type and sex on dual task for performance of paper and pencil motor tracking, the data from the single and dual tasks were entered separately into a 2 (group: control vs. ADHD) × 2 (group: girls vs. boys) × 2 (condition – type of task: single vs. dual) analysis of variance (ANOVA). For the computerised motor tracking task the ANOVA showed a significant effect of sex by factor interaction F(1,174)=4.418, *MSE*=147.574, p<.037, but the effect of factor, groups and their interaction were not significant. The effect of ADHD type was not significant for paper and pencil dual-task measures. In general in both groups, boys perform better than girls (Figure 8). The effect of comorbidity wasn't found for the paper and pencil motor tracking task.

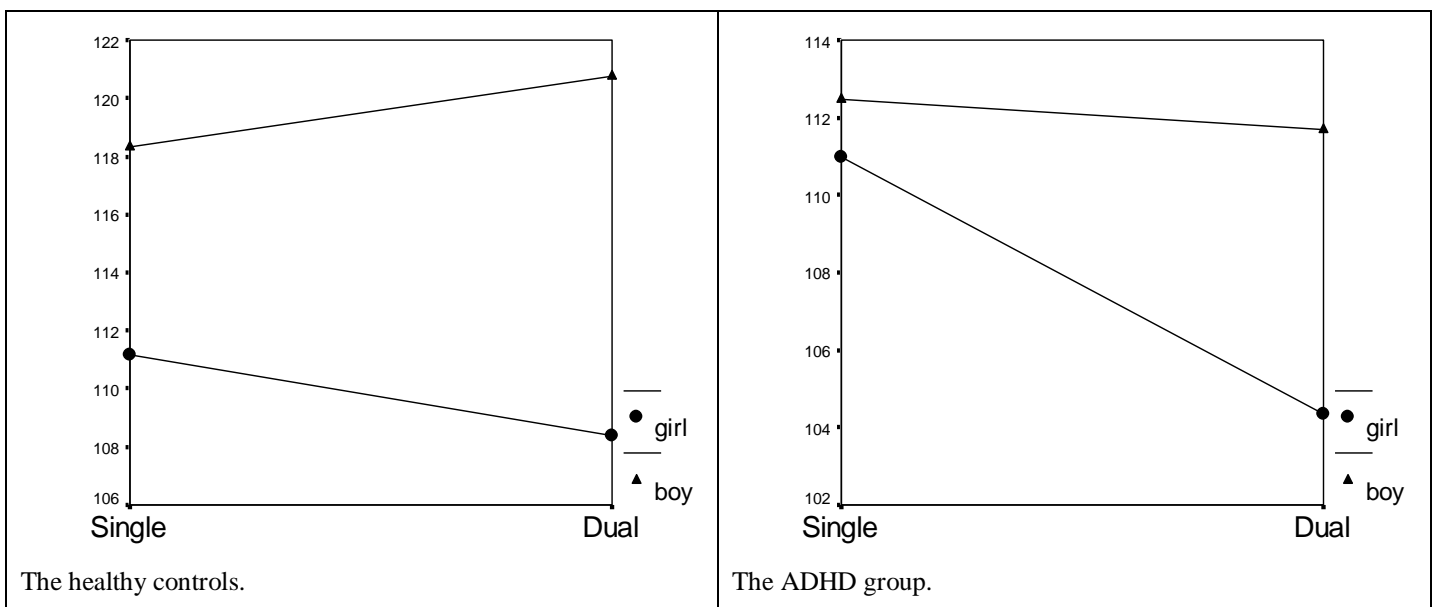

Figure 8. The mean performance of paper and pencil motor tracking task by sex for the study groups.

The sex differences were found for computerised µ scores for the healthy controls t(79.66)=2.149, p<.035 (for girls M=103.67, σ=9.17, for boys M=99.02, σ=11.13), but it was significant at one tail for the ADHD group t(89)=1.78, p<.078 (for girls M=101.37, σ=11.13, for boys M=96.76, σ=12.27). The sex differences were not significant for paper and pencil µ scores for healthy controls and ADHD group.

The effect of ADHD types and comorbidity weren't found for the List Memory Task of the paper and pencil dual-task method.

The computerized and paper and pencil µ scores were not significantly determined by the factors of ADHD types and comorbidity disorders.

**The task difficulty and dual-task performance.**

Table 7 reports the means of the 6 dual tasks (3 motor tracking and 3 digits recall tasks) for three difficulty levels (easy task, at span, difficult task) for healthy controls and children and adolescents with ADHD (Figure 9).

Table 7. Mean performances of the 6 dual tasks for three difficulty levels (easy task, at span, difficult task) for study groups.

| M (σ) | Single motor task-tracking speed at span | Single motor task - .5 proportion of tracking speed | Single motor task - 1.5 proportion of tracking speed | Dual motor task- tracking speed at span | Dual motor task - .5 proportion of tracking speed | Dual motor task - 1.5 proportion of tracking speed | Dual motor task- Digit Span-2 | Dual motor task- Digit at Span | Dual motor task- Digit Span+2 |
|---|---|---|---|---|---|---|---|---|---|
| | | | | Digit at span | | | Tracking speed at span | | |
| Control | 54.26 (8.45) | 90.23 (8.05) | 26.64 (7.93) | 57.84 (9.72) | 86.58 (14.00) | 27.00 (8.69) | 57.84 (9.72) | 59.78 (9.78) | 55.84 (11.03) |
| ADHD | 52.13 (9.13) | 88.87 (9.03) | 25.88 (7.87) | 53.67 (10.55) | 86.12 (8.75) | 25.37 (7.80) | 53.67 (10.55) | 56.24 (10.33) | 50.76 (11.64) |
| | Single verbal task- Digit at Span | Single verbal task- Digit Span-2 | Single verbal task- Digit Span+2 | Dual verbal task- Digit at Span | Dual verbal task- Digit Span-2 | Dual verbal task- Digit Span+2 | Dual verbal task- tracking speed at span | Dual verbal task- .5 proportion of tracking speed | Dual verbal task- 1.5 proportion of tracking speed |
| | | | | Tracking speed at span | | | Digit at span | | |
| Control | .82 (.12) | 1.00 (.01) | .11 (.15) | .78 (.13) | 1.00 (.01) | .16 (.19) | .78 (.13) | .74 (.15) | .79 (.16) |
| ADHD | .82 (.14) | 1.00 (.01) | .12 (.18) | .76 (.18) | 1.00 (.01) | .16 (.19) | .76 (.18) | .73 (.19) | .79 (.15) |

To investigate the task difficulty effect on ADHD and its subtypes in dual-task paradigm 6 dual-tasks (3 motor tracking and 3 digits recall tasks) were entered separately into a 2 (group) × 3 (condition – task difficulty: easy task, at span, difficult task) ANOVA.

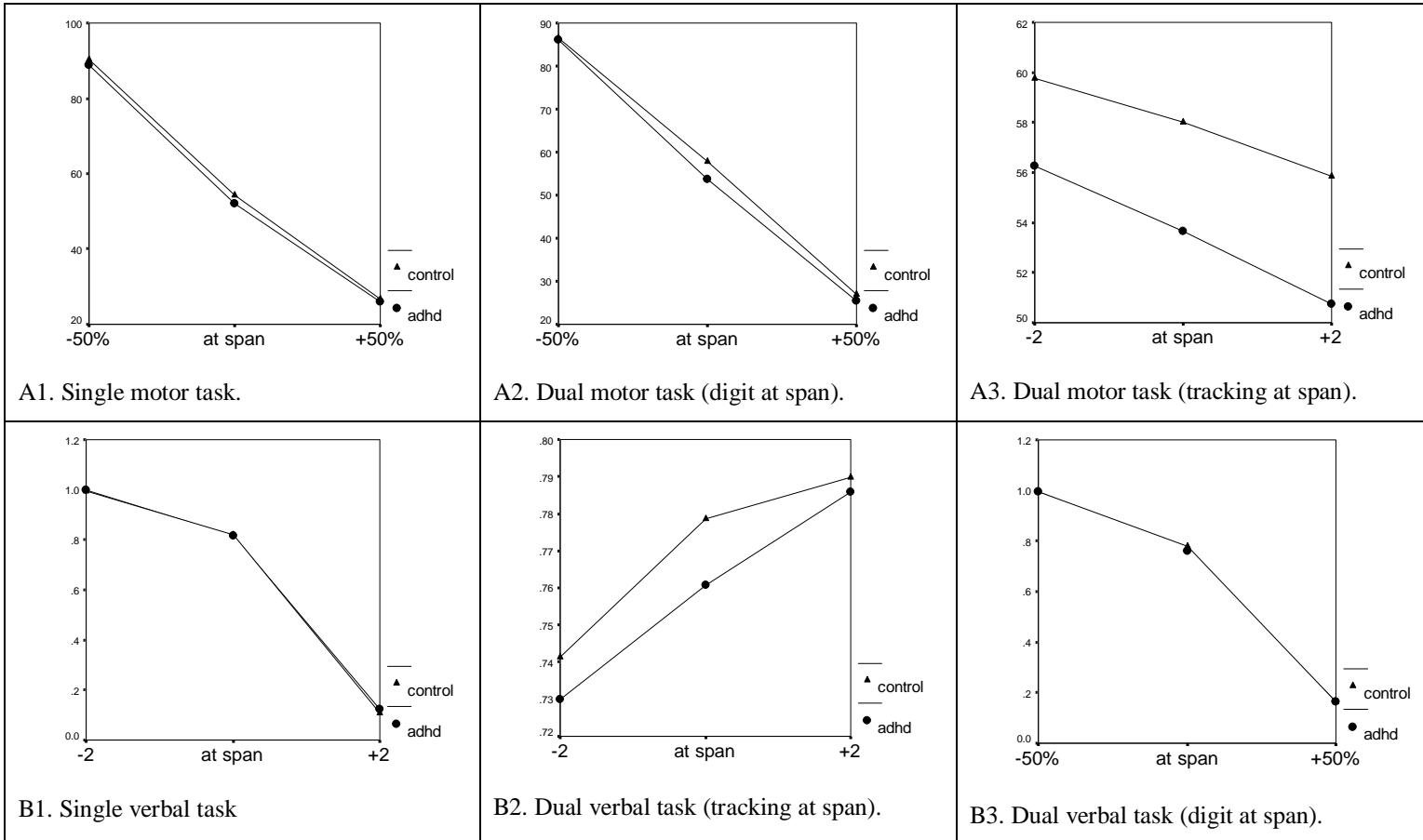

Figure 9. A. Mean percentage of time on target for motor tracking with 1. tracking performed alone and tracking demand varied (single task), 2. tracking demand varied concurrently with recall of fixed length sequences of digits (digit at span), and 3. tracking demand fixed concurrently with recall of varied length sequences of digits (tracking at span) for study groups.

B. Mean percentage of lists recalled correctly with 1. recall performed alone and sequence length varied (single task), 2. recall of varied length sequences of digits (tracking at span) concurrently with tracking demand fixed, 3. recall of fixed length sequences of digits concurrently with tracking demand varied (digit at span) for study groups.

For the tracking performed alone and tracking demand varied (single task) ANOVA showed a significant effect of type of task, $F(2,354)=4220.895$, $MSE=42.779$, $p<.0001$, but no effect of group ($F<1$) and no interaction ($F<1$). For the tracking demand varied concurrently with recall of fixed length sequences of digits (digit at span) ANOVA showed a significant effect of type of task, $F(2,354)=2479.548$, $MSE=65.301$, $p<.0001$. The effects of group $F(1,177)=3.366$, $MSE=173.825$, $p<.068$ and interaction $F(2,354)=2.434$, $MSE=65.301$, $p<.089$ were significant at one tail. For the tracking demand fixed concurrently with recall of varied length sequences of digits (tracking at span) ANOVA showed a significant effect of type of task, $F(2,356)=31.873$, $MSE=31.391$, $p<.0001$ and of group $F(1,178)=9.367$, $MSE=269.206$, $p<.003$, but the effect of interaction was not significant. For the recall performed alone and sequence length varied (single task) ANOVA showed a significant effect of type of task, $F(2,192)=3069.299$, $MSE=.01$, $p<.0001$. and of group $F(1,178)=9.367$, $MSE=269.206$, $p<.003$, but no effect of group and no interaction (F<1). For the recall of varied length sequences of digits (tracking at span) concurrently with tracking demand fixed ANOVA showed a significant effect of type of task, $F(2,356)=1900.32$, *MSE*=.02, $p<.0001$, but no effect of group and no interaction (F<1). For the recall of fixed length sequences of digits concurrently with tracking demand varied (digit at span) ANOVA showed a significant effect of type of task, $F(2,354)=14.936$, *MSE*=.008, $p<.0001$, but no effect of group and no interaction (F<1). In general the significant effect of group showed significantly better performance of healthy controls in comparison to ADHD group. But no significant interactions were found for verbal and motor task difficulties and two study groups or subtypes of ADHD.

## Discussion

It is proposed that that ADHD is associated with the disturbances in motivational processes and executive dysfunction (Sonuga-Barken EJS, 2005). The executive functions cover a wide range of cognitive processes. Part of them could be related to the neural mechanisms that are dysfunctional in ADHD.

The study found that dual-task coordination is available in children and adolescents with ADHD regardless of belonging of subject to the particular subtype of ADHD. It was shown as in the paradigm when both component tasks were titrated that was done in computerized version of the task as in the paradigm when only one task was titrated that was the case with paper and pencil version of the dual task. This result was more apparent for the paper and pencil version of the dual task. It was found that task difficulty in dual-task paradigm doesn't affect disproportionately children and adolescents with ADHD in comparison to age and years of education matched healthy controls. The ADHD subjects have no specific deficit in dual task performance regardless of their limitations in attentional processing. Thus we can suggest that controversial results found on dual-task coordination deficit in ADHD on one hand is determined more by the lack of purity of other dual-task paradigms in terms of presence in their performance the need of other cognitive and executive functions that are deficient in ADHD, and on the other hand by the task difficulty that differentially taxes the restricted cognitive resources of ADHD in comparison to the healthy controls.

According to the studies on neuroanatomical structures underlying the dual-task functioning in healthy adults was suggested that anterior cingulated area appears to be associated with dual task demands in the healthy brain, over and above what is required for each task individually (Johannsen et al., 1999). It can be proposed that dual-task coordination in children with ADHD is related to neuroanatomical mechanisms different from those compromised in ADHD. The future study of neuroanatomical correlates of dual-task demands in healthy children in comparison to children and adolescents with ADHD could find separation of neural processes underlying the dual-task performance and executive dysfunction associated with ADHD and will gain our understanding of biological bases of ADHD.

The study supported the concept of a separate cognitive function associated specifically with dual-task performance and provided evidence for the independence of the dual-task coordination function from the other executive functions that are compromised in ADHD.

The paper and pencil version of the dual task using the "Tbilisi paper and pencil motor task" proved to be successfully applicable in the educational settings. The paper and pencil version of the dual-task method gave the same results in ADHD and healthy controls as computerised version of the dual task. It can be used in school settings with children to test working memory functioning and uncover its relationships with other cognitive developmental disorders.

According to study findings the teaching strategies and rehabilitation programs for children with different subtypes of ADHD could include divided attentional demands if the individual tasks would correspond to the ability levels of children with ADHD.

The general motor functioning deficit is known to be associated to ADHD (Meyer & Sagvolden, 2006). Approximately 60% of children who are formally diagnosed as having ADHD are reported to have poor motor coordination (Henderson et al., 2007). The study showed the balance deficit in ADHD that we proposed could be associated with the fronto-cerebellum circuit dysfunction found in ADHD. The motor deficit found in ADHD affected performance on computerized motor tracking tasks in dual condition but didn't determine the overall dual-task performance. The general motor functioning only slightly affected dual-task performance for healthy controls. It can be concluded that deficits of motor functioning in ADHD don't disturb performance of dual-task coordination when the individual motor task difficulty level is adjusted in accordance to the individual ability level of subjects with ADHD.

The subtypes of ADHD don't differ on dual-task coordination ability. The presented dual-task paradigm can not be used as a criteria or tool for differential diagnostics of subtypes of ADHD. The frequently concomitant to ADHD comorbid disorders revealed differences in single and dual motor task performance on computerized version of the task in hyperactive/impulsive and inattentive subtypes of ADHD, but didn't affect the overall dual-task coordination.

The sex differences were found on computerized and paper and pencil motor tracking task performance in children and adolescents with ADHD. The girls performed computerized motor task better than boys but boys perform paper and pencil motor task better than girls in children and adolescents with ADHD. In General sex difference was found on overall dual task performance for healthy control group, but it wasn't detected for ADHD group.

The future study of dual-task coordination ability in healthy control children and children and adolescents with ADHD could concentrate on dual-task coordination strategies in ADHD to find the cognitive recourses that this group of children is using to perform this task at a normal level. It would be worthwhile to conduct dual-task neural imaging study and find responsible neuroanatomical mechanisms of dual-task performance in healthy children and ADHD group and compare it with the performance of deficient in ADHD cognitive functions that will deepen our knowledge about the neural basis of ADHD.